\documentclass{aastex6}

\begin{document}

\title{On the Rotation of Sunspots and Their Magnetic Polarity}

\author{Jianchuan Zheng\altaffilmark{1}, Zhiliang Yang\altaffilmark{1}
        and Kaiming Guo\altaffilmark{1}}
\affil{Department of Astronomy, Beijing Normal University,
          Beijing 100875, China}
\email{zhengjc@mail.bnu.edu.cn}
\email{zlyang@bnu.edu.cn}

\and

\author{Haimin Wang\altaffilmark{2, 3} and Shuo Wang\altaffilmark{2}}
\affil{Space Weather Research Laboratory, New Jersey Institute of
       Technology, University Heights, Newark, NJ 07102-1982, USA}
\affil{Big Bear Solar Observatory, New Jersey Institute of
       Technology, Big Bear City, CA 92314-9672, USA}

\begin{abstract}
The rotation of sunspots of 2 yr in two different solar
cycles is studied with the data from the Helioseismic and
Magnetic Imager on board the \it Solar Dynamics Observatory
\rm and the Michelson Doppler Imager instrument on board the
\it Solar and Heliospheric Observataory.\rm
We choose the $\alpha$ sunspot groups and the relatively large and
stable sunspots of complex active regions in our sample. In the year of 2003,
the $\alpha$ sunspot groups and the preceding sunspots tend to rotate
counterclockwise and have positive magnetic polarity in the northern
hemisphere. In the southern hemisphere, the magnetic polarity and
rotational tendency of the $\alpha$ sunspot groups and the preceding
sunspots are opposite to the northern hemisphere. The average rotational
speed of these sunspots in 2003 is about $0^{\circ}.65 \rm \ hr^{-1}$.
From 2014 January to 2015 February, the $\alpha$ sunspot groups
and the preceding sunspots tend to rotate
clockwise and have negative magnetic polarity in the northern hemisphere.
The patterns of rotation and magnetic polarity of the southern
hemisphere are also opposite to those of the
northern hemisphere. The average rotational speed of these
sunspots in 2014/2015 is about $1^{\circ}.49 \rm \ hr^{-1}$.
The rotation of the relatively large and stable preceding 
sunspots and that of the $\alpha$
sunspot groups located in the same hemisphere have opposite rotational
direction in 2003 and 2014/2015.
\end{abstract}


\section{Introduction}
The magnetic field, to which almost all solar active phenomena
are related, plays an important role in solar physics.
Sunspots are the most obvious characteristics of the magnetic field on
the Sun \citep{tho2008}. Although the earliest observation of sunspots
with telescopes dates back to the time of Galileo, the magnetic
field was found in 1908 by George Hale, who observed line splitting
and polarization in sunspot spectra by solar tower at the Mount
Wilson Observatory \citep{hal1908}. The 11 yr activity cycle, based
on the variation of the sunspot numbers and known since the
mid-nineteenth century \citep{sch1844}, is a prominent feature of the Sun.
Newly emerged sunspots in an early solar cycle are located on
midlatitudes of the Sun, and successively appear closer to the
solar equator as the cycle progresses. The
latitudinal distribution of sunspots and its progression over the
sunspot cycle follow a ¡°butterfly diagram¡±. \citet{hal1919} further
studied the polarity and distribution of sunspots and established Hale's
laws. Sunspots generally appear as pairs which contain a preceding sunspot
and a following sunspot. The preceding and following sunspots have opposite
magnetic polarity, and each polarity is reversed in the two hemispheres
(Hale's law). In about 11 yr, the preceding sunspots in one
hemisphere (e.g. the northern hemisphere) keep one magnetic polarity,
in the next 11 yr, preceding sunspots switch to opposite
magnetic polarity. So the magnetic cycle of sunspots is about 22 yr
\citep{hal1925}, which is also known as the Hale Cycle. Although the periodic
change of sunspots is well observed, our understanding of its
physical nature is still evolving.


The twist, shear, linking, and kinking of magnetic field are often
described by magnetic helicity. Helicity was studied for more than 20 yr
\citep{pev2008}. Some statistical studies of active regions showed
that the helicity tended to be negative in the northern hemisphere and
positive in the southern hemisphere, which is known as the hemispheric
helicity rule \citep{pev2014}. The hemispheric preference is
statistically weak for active regions
(e.g. \citealt{see1990, pev1995, pev2001, pevh2008, abr1997,
bao1998, hag2004, hag2005, zha2006, hao2011, liu2014}), from
$\sim 58\%$ to $82\%$, and it does not vary with the solar cycle.


Besides the magnetic field, rotation seems to be the prominent property
of sunspots.
The rotational motions exist on the solar surface and atmosphere
ubiquitously and constantly (see, e.g. \citealt{kom2007, att2009, zha2011,
wed2012}). Sunspot rotation was first observed by \citet{eve1910} a
century ago. Recently, more and more rotating sunspots have been identified.
\citet{yan2012} gave the definition of rotating sunspots: a sunspot can be
regarded as a rotating sunspot when it rotates around its umbral center or it
rotates around another sunspot within the same active region. Subclasses of
six rotational types were given by \citet{yanx2008}.
\citet{zha2007} reported a large rotational angle of sunspots up to
$240^{\circ}$. \citet{min2009} found that a small sunspot of positive
polarity in AR 10930 rotated counterclockwise about its center by
$540^{\circ}$ within 5 days using a nonlinear affine velocity estimator.
\citet{yan2009} also found this sunspot rotating $259^{\circ}$
over 4 days before an X3.4 flare.
\citet{bro2003} studied statistically seven sunspots using white-light
\it Transition Region and Coronal Explorer (TRACE) \rm data, and found
that they rotated between $40^{\circ}$ and
$200^{\circ}$ within 3-5 days. Besides, \citet{yan2008} found 182
rotating sunspots among 2959 active regions from 1996 December to 2007 December. The number
of rotating sunspots with positive polarity was more than the number of
those with negative polarity in the northern hemisphere, while the pattern
was opposite in the southern hemisphere. \citet{zhu2012} studied the
relationship between rotating sunspots and emergence of magnetic flux tubes
and found that the rotational angular velocities of sunspots in active regions
with flux emergence were larger than those without flux emergence.
Theoretically, \citet{stu2015} simulated
photospheric footprints or sunspots of the flux tube and found that they
are undergoing rotation.


Our study is motivated by the following questions about
the rotational motion of sunspots: Why do some sunspots rotate?
Is there any relation between the rotation and magnetic field
of sunspots? Is there periodicity about rotation of sunspots?
To seek answers to these questions, we use the data of two solar
cycles, the cycle 23 from 1997 to 2008 and the cycle 24 from 2008 to
2019, to obtain the statistical results. The paper is organized
as follows. The data we use are described in section \ref{data}.
The method of analysis of the rotation of sunspots is presented in
section \ref{method}. Section \ref{analysis} shows data processing. We
summarize and discuss the results in section \ref{summary}.


\section{Observational Data}\label{data}
The data used in our study are taken from the Michelson Doppler Imager (MDI) on board
the \it Solar and Heliospheric Observatory (SOHO) \rm spacecraft \citep{sch1995}
and the Helioseismic and Magnetic Imager (HMI; \citealt{sch2012}) on board
the \it Solar Dynamics Observatory (SDO). \rm


The MDI made observations in the Ni I line at 6768 $\mathrm\AA$;
longitudinal
magnetograms were constructed by measuring the Doppler shift of the
line separately with right and left circularly polarized wave plates.
The difference between these two is a measure of the Zeeman splitting
and is roughly proportional to the magnetic flux density, the
line-of-sight (LOS) component of the magnetic field averaged over
the resolution
element. The $1024\times1024$ pixel continuum and magnetogram images
have pixel resolution of $2''$ .


The spatial resolution of HMI is $0''.5$ pixel$^{-1}$ and the spectral
line is at Fe I 6173 $\mathrm \AA$. We use the
continuum and magnetogram images of HMI to study rotation of sunspots.
Continuum intensity and LOS magnetic field strength
are two of the LOS observables that were obtained by reconstructing the
spectral line profile from the measurements at the six positions
\citep{cou2012}.


Basically, the sunspots under study can be divided into two classes:
the $\alpha$ sunspot groups and the complex sunspot groups. The magnetic
structures of the $\alpha$ sunspot groups are relatively simple. The
complex sunspot groups usually have preceding sunspots and following
sunspots. The sunspots within each group are interacting with each other. The
eruptive events like coronal mass ejections (CMEs) and flares occur much more
frequently in the complex active regions.
We select the sunspots with the following principles:
First, sunspots with radius larger than about
$8\ \rm Mm$ (about $10''.0$) and the $\alpha$
sunspot groups with simple magnetic structure are chosen.
Perhaps limited by spatial resolution, very tiny sunspots are not showing the
rotational properties obviously. Large sunspots show
rotational properties at a different scale. Second, their structure
should be relatively stable. There is no interaction
with other sunspots, and they keep their features for several
days. Next, we choose the sunspots close to the center of the solar
disk from about
E30 to W30.
The locations of sunspots are obtained from the
website (http://www.solarmonitor.org/).
When the sunspots are roughly located on the center of the
solar disk, we record the maximum magnetic field from the data.
The maximum strength of sunspots has a value on the order of
a kilogauss within several days in our study.


The data are obtained through the Joint Science Operations Center (JSOC)
Web server (http://jsoc.stanford.edu/). The intensitygrams of MDI are
obtained with the MDI Full Disk Intensity Continuum, and the magnetograms of
MDI are obtained with the MDI Full Disk 96 m Magnetogram at a cadence of 96 minutes.
The data of sunspots are from 2003 January to 2003 December, which contains
5 $\alpha$ sunspots group and 37 preceding sunspots of complex active regions.
The intensitygrams and magnetograms of HMI are obtained from HMI Active Region
patches (HARPs) at a cadence of 12 minutes. The data of these sunspots are from
2014 January to 2015 February, which contain 18 $\alpha$ sunspot groups and
6 preceding sunspots and 10 following sunspots of complex active regions.


\section{The Analysis Methodology}\label{method}
\subsection{Time-slice Method}\label{submethod}
First, from the images of the photosphere, the regions of interest within
active regions are chosen. The image that is close
to the center of the solar disk is used as a reference image, because of
its smaller projection effect. We align the time-sequence continuum
intensitygram and LOS magnetogram data to the reference
image with the `rot\_xy' procedure, which takes into account the solar
differential rotation effect. After repeating this procedure, we then
create a movie from these time-series images
to make sure that the regions are in the frame. The
movie can also be used to visualize rotation of sunspots.


The $\alpha$ sunspot group has a simple structure, but
complex active regions contain sunspots with different polarities.
Thus, analysis of individual sunspots is needed.


In order to derive the rotational parameters of the sunspots, we
follow the procedure developed by \citet{bro2003}. It is also used by many
other authors (e.g., \citealt{bro2003, jia2012, lia2015}). As sunspots on the
photosphere move, merge, deform, and disappear, we follow the center
of umbra. There are two ways to determine the center of the sunspot:
(1) the point of minimum intensity
in the white-light images, and
(2) the point with absolute maximum value in the magnetograms. In our
images, we choose the points with maximum value of the absolute
magnetic field.


We uncurl the circular region of the sunspot in the white-light
image by transforming
from the Cartesian ($x-y$) frame to the polar ($r-\theta$) frame,
that is, cutting off the circular region of the sunspot and unfolding it
into the $r-\theta$ plane. In this process, the original location to
cut off is parallel to the $x$ axis, and then we uncurl it counterclockwise from
$0^{\circ}$ to $360^{\circ}$. Then we get the intensitygram of the
sunspot in the $r-\theta$ frame. Following this process on all the
time-series images, one can obtain the $\theta$ variation of characters
as a function of time. We can extract one line at each time
by fixing the radius on the
$r-\theta$ plane. The time series lines will be formed by the intensitygrams at
the circle with fixed radius $r_c$ in the time$-\theta$ plane, and the
rotational trend can be found.


The radius $r_c$ is chosen with the following principles: when there
are enough characteristic points to show the trend, $r_c$ is near and
in the umbra of the sunspot. The reason is discussed in section
\ref{r_chosen}. The rotational characters of sunspots
can be obtained, and angular velocity can be evaluated.

\subsection{Radius Chosen in Time-slice Method}\label{r_chosen}
About the radius chosen for the rotational sunspots, \citet{bro2003}
suggested that the penumbra is rotationally feature-rich, and thus
feature-tracking results can be obtained easily.
easy to trace the rotational features. Close to the umbra, there is a less prominent
rotational feature to trace. However, as the motion of the penumbra is
complicated and the structure of the penumbra changes more rapidly, there
are some uncertainties in the determination of rotation. But the umbra is
relatively stable and may present a global rotating structure. So, we
analyze the different parts (umbra and penumbra) of sunspots to
determine which one is better for obtaining the real rotation.
In the following, we present two examples to show how to analyze
the sunspot rotation.

\subsubsection{AR 11158 -- a Complex Active Region (Evolving Sunspots)}
\begin{figure}
 \centering
  \includegraphics[width=0.9\textwidth]{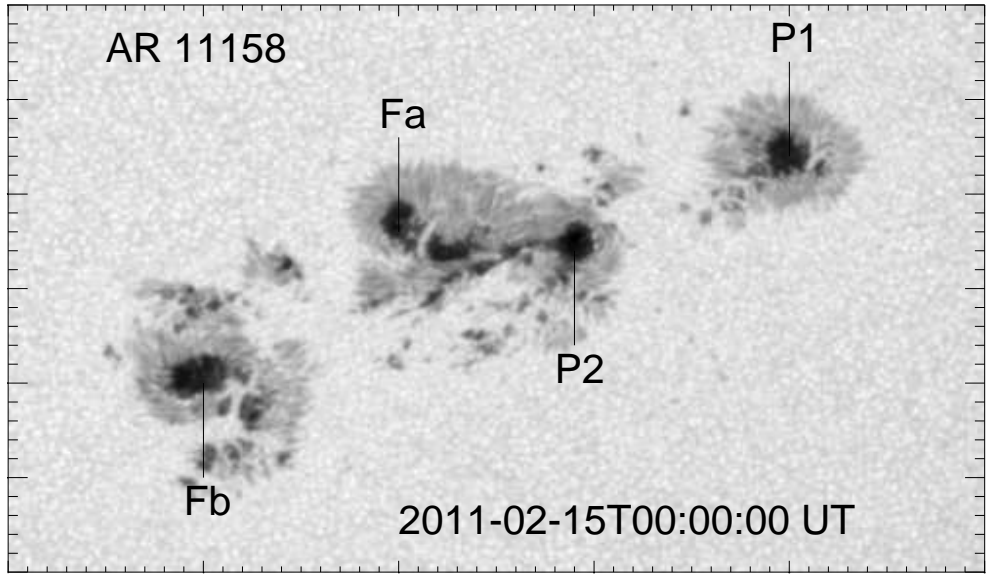}
  \caption{The HMI continuum intensity image of AR 11158 on 2011 February
           15. The AR consists of four main sunspots, ``P1", ``P2", ``Fa" and
           ``Fb". ``P1" and ``P2" have positive magnetic polarity, and ``Fa"
           and ``Fb" have negative magnetic polarity.}
  \label{img11158_single}
 \end{figure}

NOAA AR 11158 was first emerged on 2011 February 11 and grew rapidly
and developed into to a $\beta\gamma\delta$-type active region on 2011 February
13. The AR consists of four main sunspots: `P1' and `P2' with positive
magnetic polarity, and `Fa' and `Fb' with negative magnetic polarity, which is
shown in Figure \ref{img11158_single}. The data are obtained
from HMI/\it SDO. \rm At the
begining of February 14, the four sunspots are forming. On February 15 and
16, these four sunspots can be distinguished clearly.

\begin{figure}
 \centering
  \includegraphics[width=0.9\textwidth]{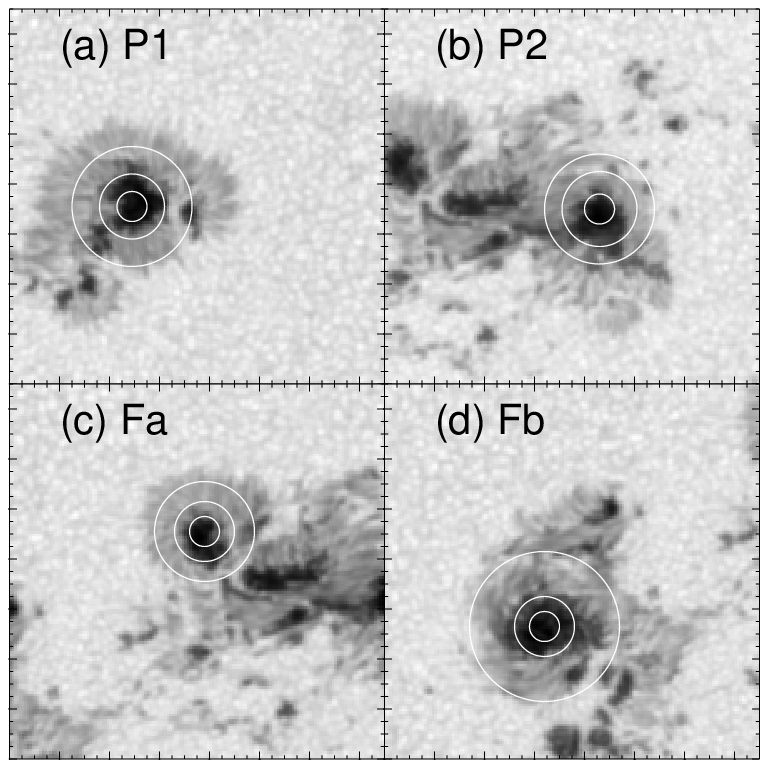}
  \caption{The four sunspots (``P1", ``P2", ``Fa" and ``Fb") in AR 11158. The
           white circles in the sunspots are the radii to produce time-slice
           figures.}
  \label{fixed_radii}
 \end{figure}

We analyze the four sunspots individually. The radii are fixed in inner
umbra, outer umbra, and penumbra to produce time slices. Figure
\ref{fixed_radii} shows the the fixed radii of time slices.

\begin{figure}
 \centering
  \includegraphics[width=0.9\textwidth]{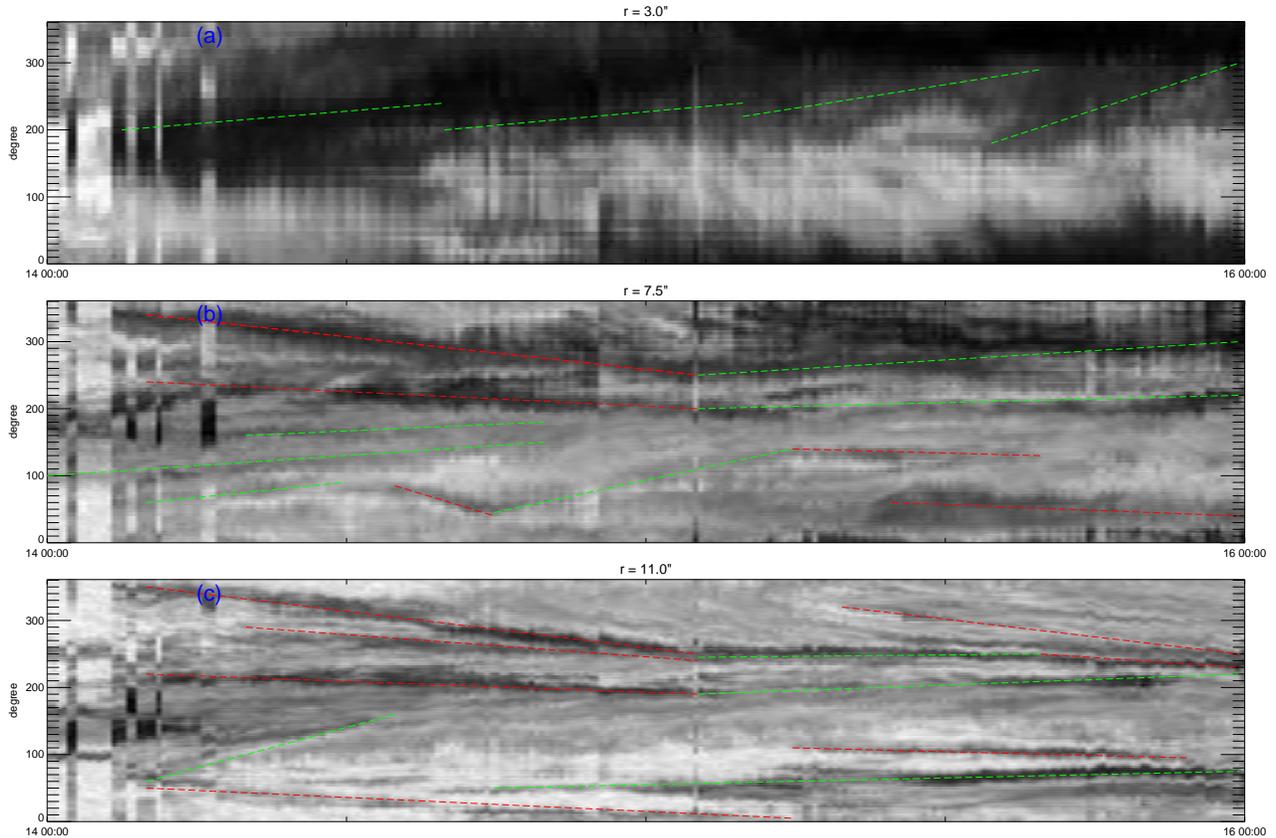}
  \caption{Time slices at different radii of P2 in AR 11158. (a)
           $r = 3''.0$, the time slice is taken from the inner umbra, and
           the green lines mark the rotational tendency
           of the sunspot. The green lines show counterclockwise
           uniform rotation of the sunspot. (b) $r = 7''.5$, the time slice
           is taken from the outer edge of the umbra. The colored lines
           mark the rotational tendency of the sunspot. There are clockwise
           and counterclockwise rotations at the same time. (c) $r = 11''.0$,
           the time slice is taken from the penumbra with HMI data. The situation
           is the same as panel (b). So panel (a) reflects the real
           rotation of the sunspot.}
  \label{abc11158_p2}
 \end{figure}

\begin{figure}
 \centering
  \includegraphics[width=0.9\textwidth]{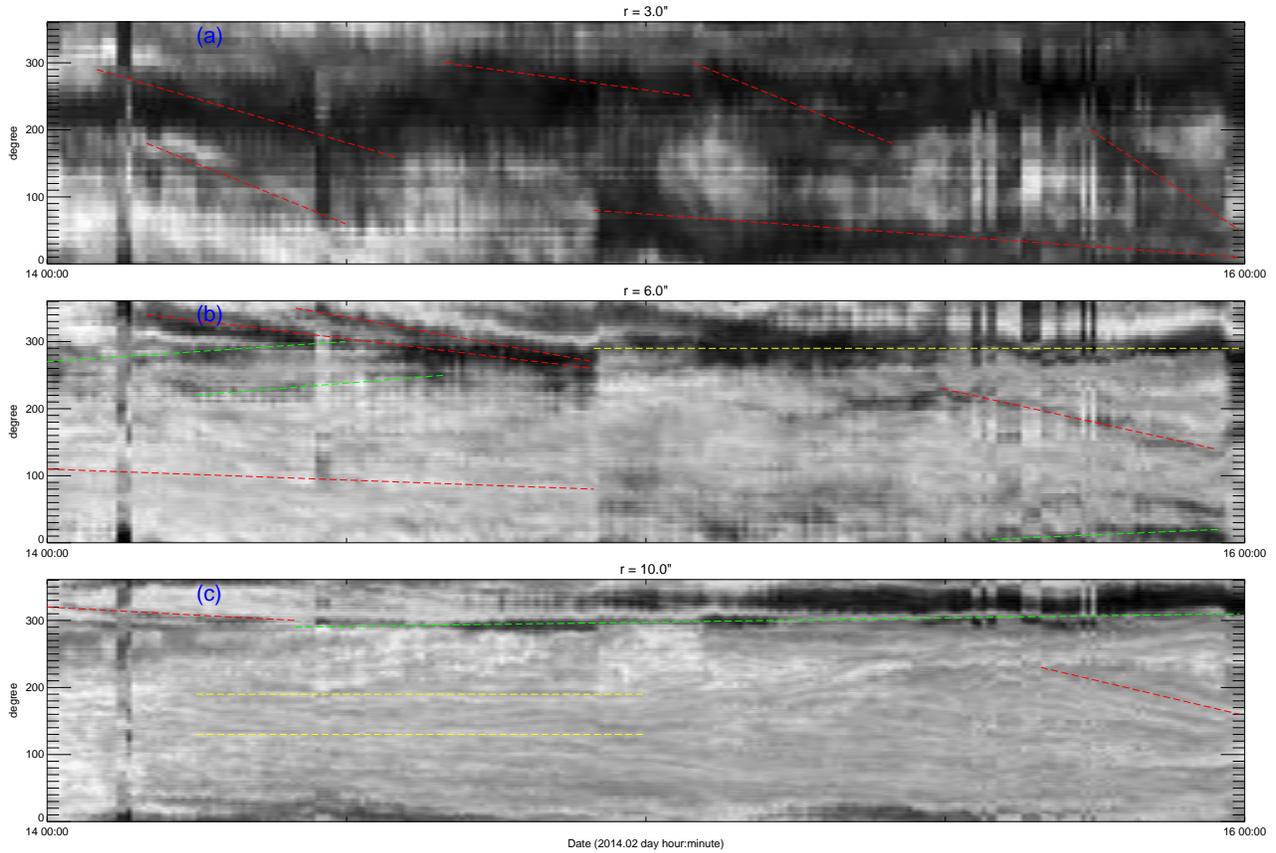}
  \caption{Time slices at different radii of Fa in AR 11158. (a)
           $r = 3''.5$, the time slice is taken from the inner umbra, and the
           red lines mark the rotational tendency of the sunspot. The sunspot
           rotates clockwise. (b) $r = 6''.0$, the time slice is taken from the
           outer edge of the umbra. The rotational clockwise and counterclockwise
           directions both exist in the entire time. (c) $r = 10''.0$,
           the time slice is taken from the penumbra. The situation is the same
           as in panel (b). The time-slice umbra may be more suitable for obtaining the
           rotation of the sunspot.}
  \label{abc11158_fa}
 \end{figure}

In the time-slice figures, Figure \ref{abc11158_p1} (a), $r = 3''.0$; Figure
\ref{abc11158_p2} (a), $r = 3''.0$; Figure \ref{abc11158_fa} (a),
$r = 3''.0$; Figure \ref{abc11158_fb} (a), $r = 3''.5$, they are all
located in the umbrae. Diagonal lines in the time slice images show
clear pattern of rotation.
When the radii are
located in the outer edge of umbrae, as in Figure \ref{abc11158_p1}
(b),  $r = 6''.5$; Figure \ref{abc11158_p2} (b), $r = 7''.5$; Figure
\ref{abc11158_fa} (b), $r = 6''.0$; Figure \ref{abc11158_fb} (b),
$r = 6''.0$, the trends are complicated. Some points
show clockwise motion and other
counterclockwise. When the radii are located in the regions of penumbrae,
as in Figure \ref{abc11158_p1} (c),  $r = 12''.0$; Figure \ref{abc11158_p2}
(c), $r = 11''.0$; Figure \ref{abc11158_fa} (c), $r = 10''.0$; Figure
\ref{abc11158_fb} (c), $r = 12''.0$, the situations are the same as in the outer edge
of the umbrae. The trends are even more complicated and it is hard to
determine the real rotational directions of the sunspots.

\begin{figure}
 \centering
  \includegraphics[width=0.9\textwidth]{abc11158_p1.eps}
  \caption{Time slices at different radii of P1 in AR 11158. (a)
           $r = 3''.0$, the time slice is taken from the inner umbra, and the green
           lines mark the rotational tendency
           of the sunspot. The green lines show that this sunspot rotates
           counterclockwise. The sunspot alway rotates counterclockwise
           on the time period. (b) $r = 6''.5$, the time
           slice is taken from the outer edge of the umbra. There are clockwise and
           counterclockwise rotational tendencies at the same time. (c)
           $r = 12''.0$, the time slice is taken from the
           penumbra. The situation is the same as in panel (b).}
  \label{abc11158_p1}
 \end{figure}

\begin{figure}
 \centering
  \includegraphics[width=0.9\textwidth]{abc11158_fb.eps}
  \caption{Time slices at different radii of Fb in AR 11158. (a)
           $r = 3''.5$, the time slice is taken from the inner umbra, and the
           red lines mark the rotational tendency of the sunspot.
           At the begining, there are some black features that show clockwise
           rotation. On February 14, there are some explosions. There
           are some clockwise (black streaks) and counterclockwise (white
           streaks) rotational features. (b) $r = 6''.0$, the time slice
           is taken from the outer edge of the umbra, and there are clockwise and
           counterclockwise rotations at the same time. It is hard to
           determine the real rotation of the the sunspot. (c) $r = 12''.0$,
           the time slice is taken from penumbra. There are some clockwise
           and counterclockwise rotation features, but not obvious. So
           panel (a) may reflect the real rotation of the sunspot.}
  \label{abc11158_fb}
 \end{figure}


The green lines of Figure \ref{abc11158_p1} (a) show sunspot P1, which rotates
counterclockwise. They all show uniform rotational tendency at
different areas and on different time periods (from 14 00:00 UT to 16
00:00 UT). However, there are streaks showing different rotational tendency
in Figure \ref{abc11158_p1} (b). The red lines represent the clockwise
rotation. The green lines mark the counterclockwise rotational features,
and the yellow lines mark the nonrotational features. Different rotational
tendencies (lines with different color) exist at the same time, as shown
in the figure. This indicates that the motion of the outer
umbra is complicated. It is not easy to determine the rotational
direction from this part. In the region of the penumbra, the situation
is the same as the outer umbra, which is shown in Figure \ref{abc11158_p1}
(c). There are lines with different colors in the same time period. This
means that there are features with different rotational patterns at the
same time. So we cannot obtain the uniform rotational tendency of the sunspot
from the region.


In Figure \ref{abc11158_p2} (a), the green lines show counterclockwise
uniform rotation of sunspot P2. The red and green lines in (b)
and (c) show that there are clockwise and counterclockwise rotations in the
same time period.  The rotational tendency in (b) and (c) is complicated and
cannot be determined.  The rotational features in the inner umbra may reflect
the real rotation of the sunspot.


Sunspot Fa mainly rotates clockwise in the umbra, which is shown in Figure
\ref{abc11158_fa} (a) with red lines. The black streaks in the figure
mainly have the tendency to move down (i.e., rotate clockwise). The red
lines, green lines, and yellow lines in (b) and (c) show that there are
clockwise and counterclockwise rotations (and nonrotating parts) at the same time.
It means that the motion in the large radius (penumbra) is complicated. The
rotational tendency in the penumbra may not be the same as the rotational
tendency in the umbra. The analysis in the umbra may be suitable for obtaining rotation
of the sunspot.


In Figure \ref{abc11158_fb} (a), the red lines show that sunspot Fb rotates
clockwise. At the begining (before February 14), there are some features
(black streaks) that show clockwise rotation. Starting from February 14, many
C class flares occurred. The motion in this area is much larger. In the time-slice
figure, there are some black streaks showing clockwise rotation,
and there are also some white streaks showing counterclockwise rotation.
After February 14, there are also some black streaks showing clockwise
rotational tendency. However, there are clockwise and counterclockwise
rotations at the same time in Figure \ref{abc11158_fb} (b) and (c). It is
hard to determine the rotational tendency of the the sunspot. In Figure
\ref{abc11158_fb} (c), there are some clockwise and counterclockwise
rotation features, but not obvious. According to the above
discussion,
we find that the umbra rotates clockwise. The motion in the penumbra mainly comes
from the consequence of the flares. It may be different from the rotation
of the umbra.


From these time-slice figures, we can see that the rotational tendency is
uniform at small radius (inner umbra) with black streaks.
However, at large radius (penumbra), the rotational
tendency is complicated. Some points show clockwise rotation, but other points
show counterclockwise rotation at the same time. For the complicated sunspots,
we suggest that the small radius (in the umbra) shows the true rotation. The
large-radius cases may contain the flow motion of the penumbra. It may not be the
true rotation of the sunspots.

\subsubsection{AR 12005 -- $\alpha$ Sunspot Group (Relatively Stable Sunspot)}
\begin{figure}
 \centering
  \includegraphics[width=0.6\textwidth]{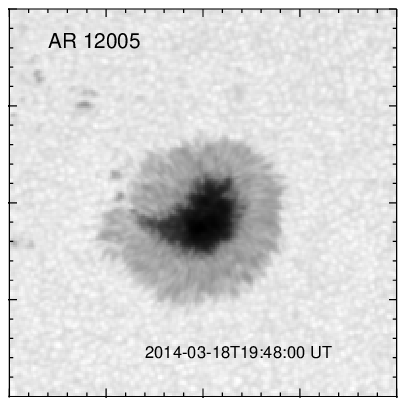}
  \caption{HMI continuum intensity image of AR 12005 on 2014 March
           18. The AR is an $\alpha$-type sunspot group with negative
           magnetic polarity.}
  \label{img12005_single}
 \end{figure}

NOAA AR 12005 is always an $\alpha$-type active region from 2014 March 16 to
March 20 (Figure \ref{img12005_single}). Its structure changes very little during this
period. The rotational tendency are shown in Figure \ref{abc12005}. When
$r = 4''.0$ (Figure \ref{abc12005} (a)), the rotational features traced are
located in inner umbra. The red line shows that the sunspot rotates
clockwise. The rotation of larger radii (penumbra) are also analyzed (Figure
\ref{abc12005} (b) and (c)), their radii are $r = 10''.0$ and $r = 15''.0$,
respectively), and the rotational trend is the same as those at small radii. So the
rotational tendencies on the small and large radii are
always the same and show that this sunspot rotates clockwise.

\begin{figure}
 \centering
  \includegraphics[width=0.9\textwidth]{abc12005.eps}
  \caption{Time slices at different radii of AR 12005. (a)
           $r = 4''.0$, the time slice is taken from the inner umbra, and
           the red line marks the rotational tendency of the
           sunspot. (b) $r = 10''.0$ the time slice is taken from the outer edge
           of the umbra. (c) $r = 15''.0$ the time slice is taken from penumbra.
           The regimes all show the clockwise rotation of the sunspot.
           The rotational tendency can be obtained from the umbra or penumbra.}
  \label{abc12005}
 \end{figure}


For the $\alpha$ sunspot groups (the relatively stable sunspots), the
rotational trends can be obtained from the umbra or penumbra. According to
the above discussion, we choose the radius $r_c$ to be near or inside the
umbra for the complicated sunspots.


\section{Data Analysis Result}\label{analysis}
With the aligned data, we use the method which is described in section \ref{method}
to analyze the rotation of sunspots.


\subsection{A Case Study: AR 12061}\label{case}
NOAA AR 12061 is located in the southern hemisphere (S24E02) and its magnetic
polarity is positive.
Figure \ref{AR12061} (a) is the HMI intensity map
of AR 12061, and the LOS magnetic field contours are overlaid (red
represents positive polarity).

\begin{figure}
 \centering
  \includegraphics[width=0.9\textwidth]{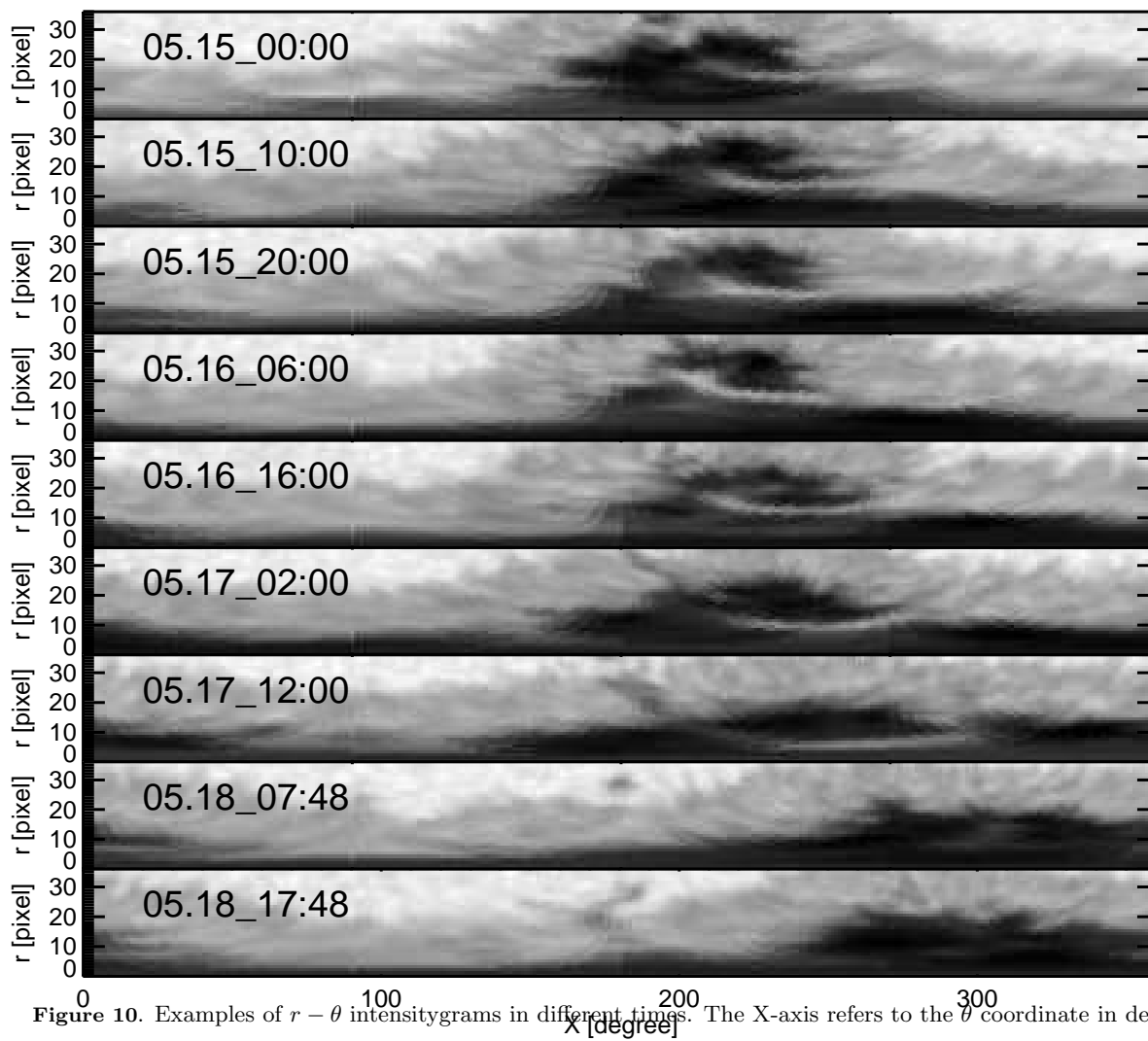}
  \caption{Examples of $r-\theta$ intensitygrams in different times.
            The X-axis refers to the $\theta$ coordinate in
           degrees.}
  \label{uncurl12061}
 \end{figure}

After choosing the point of maximum intensity of the magnetograms as
the center of the circle with a radius of 35 pixels ($17''.5$), shown
in Figure \ref{AR12061} (b), we uncurl the whole sunspot from the
Cartesian frame to the polar frame. Some images at different times on
the $r-\theta$ plane are shown in Figure \ref{uncurl12061}. From Figure
\ref{uncurl12061}, some features varying with time can be
seen, indicating that the sunspot was rotating. To display the rotational
properties better, we select the circles with radii ranging from 3 pixels
($1''.5$) to 32 pixels ($16''.0$) (Figure \ref{AR12061} b) to obtain
time slices.

\begin{figure}
 \centering
  \includegraphics[width=0.9\textwidth]{AR12061.eps}
  \caption{(a) HMI intensity image of AR 12061 overlaid by LOS magnetogram
           contour (in Gs). (b) Region to be uncurled.
           The uncurling starts at the
           westward direction and proceeds counterclockwise about the spot.}
  \label{AR12061}
 \end{figure}

\begin{figure}
 \centering
  \includegraphics[width=0.9\textwidth]{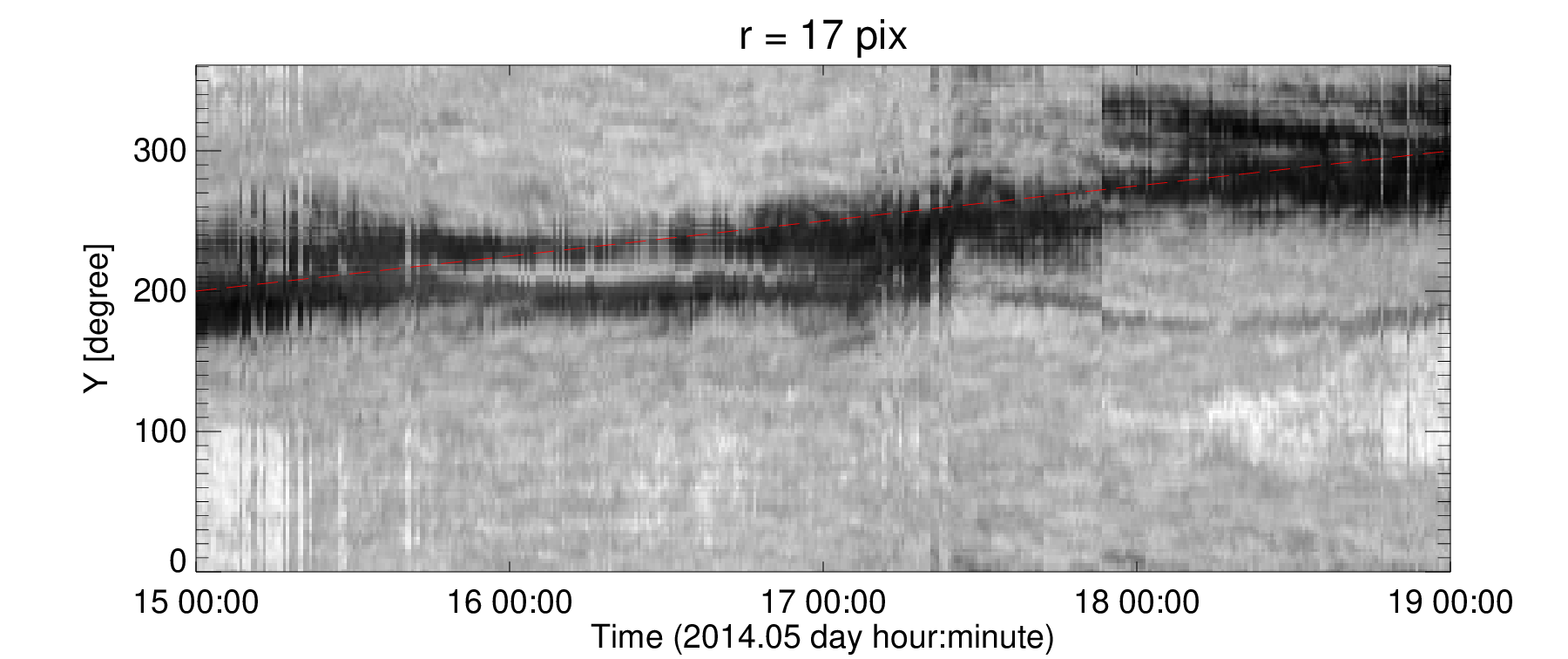}
  \caption{Time slice of AR 12061 taken from the $r-\theta$ plot at
           $r\ =\ 8''.5$. Rotation of the features can be seen by some
           diagonal streaks. We display the features as a red dashed line.
            The X-axis refers to the observational time period.
           The Y-axis refers to the $\theta$ coordinate in degrees.}
  \label{trend12061}
 \end{figure}


Figure \ref{trend12061} shows the time slices at $r=8''.5$ of Figure
\ref{uncurl12061}, which is taken from 4 days of data that contain
432 intensitygrams with a cadence of 12 minutes. There are some bright
and dark streaks in the figure. The obvious dark streak, which has
a tendency to move up, reveals the rotation of the sunspot.
This indicates that AR 12061 rotates counterclockwise.
The red dashed line indicates the trend
shown in Figure \ref{uncurl12061}. The sunspot rotates about $100^{\circ}$
within 4 days. In other words, the angular speed is about
$1^{\circ}.04\rm\ h^{-1}$. The rotational tendency is checked by comparing with
the time-sequence image movies carefully. As \citet{lia2015} pointed
out, this method did not measure the average rotation of sunspot but
obtained the rotation of the specific feature traced. However, we can get
the rotational directions of sunspots and analyze their evolution.


\subsection{The Polarity and Rotation Relationship}\label{relation}
In order to obtain the relation between the rotation and magnetic
polarity, we deal with sunspots from 2014 January to 2015 February,
as in section \ref{case}
and use the time slices of the sunspots to find the
trends of the sunspots' rotation.

\begin{figure}
 \centering
  \includegraphics[width=0.9\textwidth]{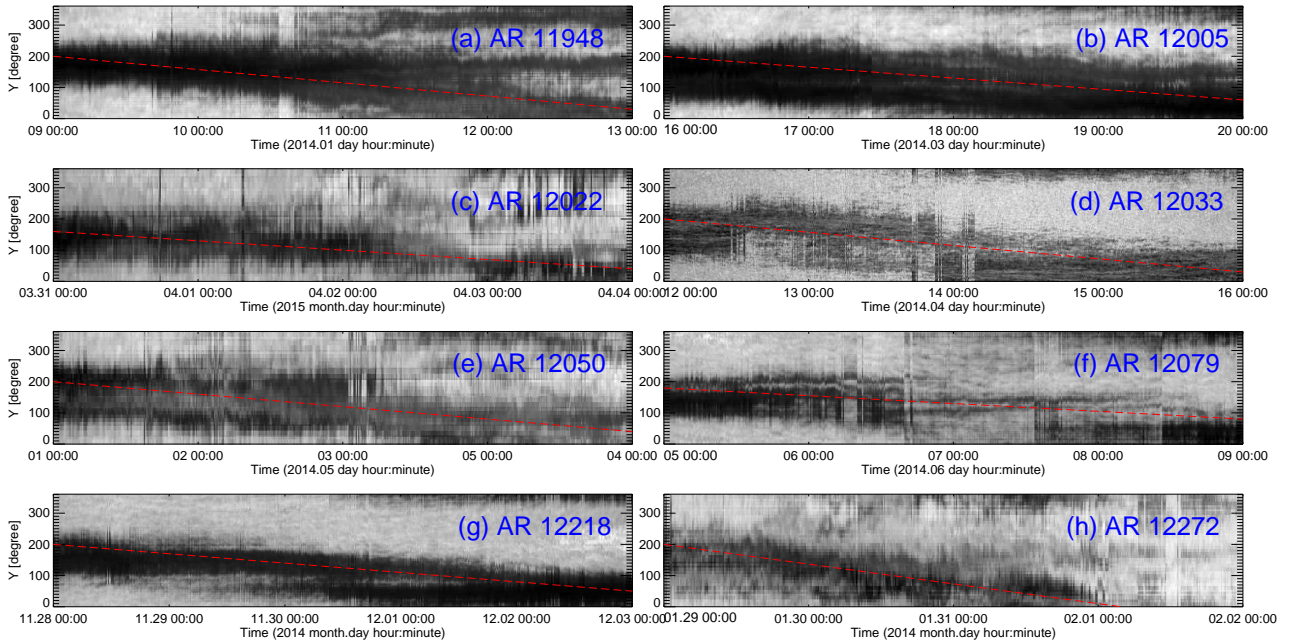}
  \caption{8 $\alpha$ sunspot groups (from (a) to (h)) are located
           in the northern hemisphere in the year of 2014/2015. Their magnetic
           polarity is negative. They rotate clockwise, and the tendency is marked
           as red dashed lines.}
  \label{trend_clockwise}
 \end{figure}
\begin{figure}
 \centering
  \includegraphics[width=0.9\textwidth]{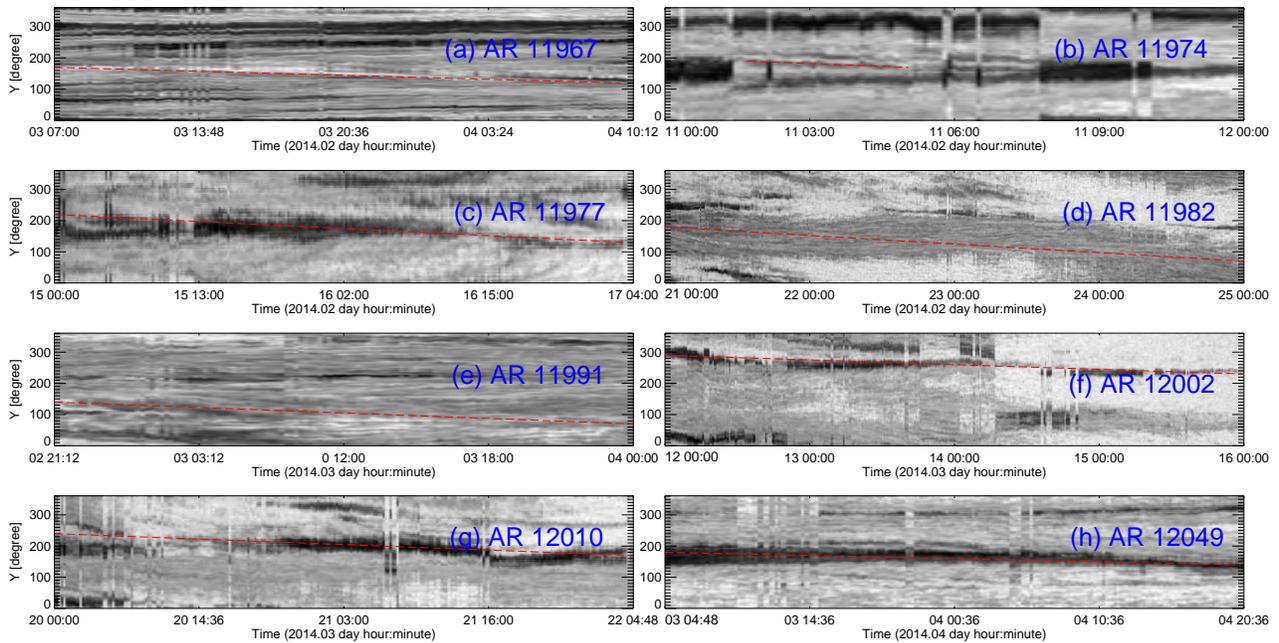}
  \caption{8 following sunspots of 2014 located in the southern
            hemisphere  rotate clockwise (marked with red dashed lines), and
            their polarity is negative.}
  \label{trend_clockwise_anormal}
 \end{figure}
\begin{figure}
 \centering
  \includegraphics[width=0.9\textwidth]{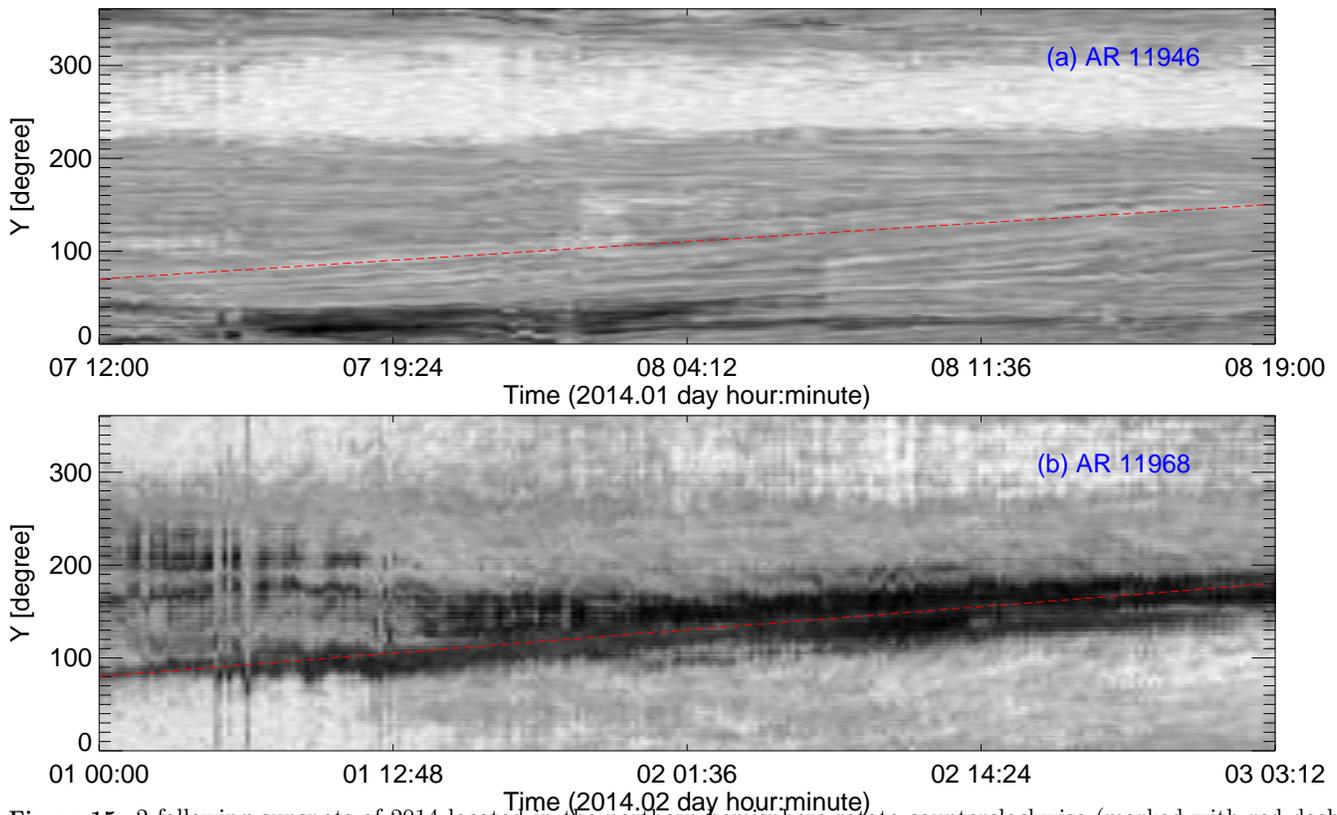}
  \caption{2 following sunspots of 2014 located in the northern
           hemisphere rotate counterclockwise (marked with red dashed lines),
           and their polarity is positive.}
  \label{trend_anormal}
 \end{figure}
\begin{figure}
 \centering
  \includegraphics[width=0.9\textwidth]{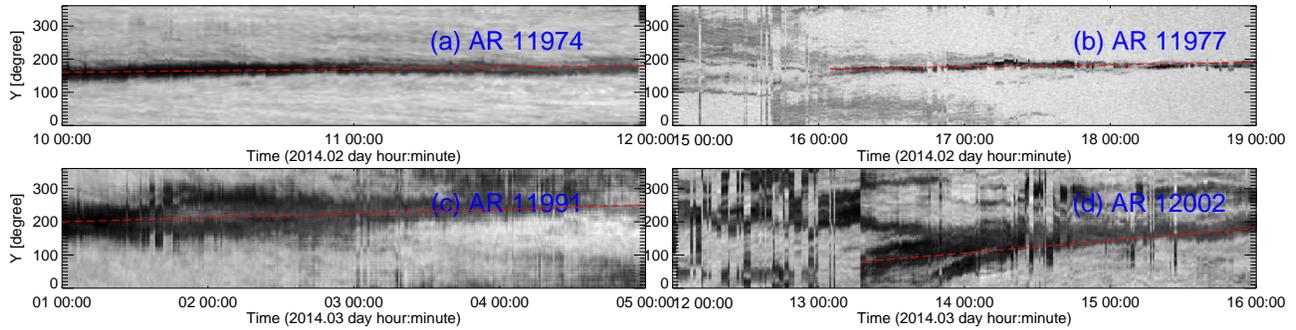}
  \caption{4 preceding sunspots of 2014 located in the southern
           hemisphere rotate counterclockwise, and their polarity is positive.
           The trend is the same as the $\alpha$ sunspot groups
           in the same hemisphere.}
  \label{preceding_south}
 \end{figure}

The sunspots' rotational directions
and angular speeds can be obtained from the trends.
Figure \ref{trend} and Figure \ref{trend_clockwise} show the
counterclockwise and clockwise rotation of the $\alpha$ sunspot
groups, respectively. The sunspots
of Figure \ref{trend} are located in the southern hemisphere, and the
sunspots of Figure \ref{trend_clockwise} are located in the northern
hemisphere. The red dashed line in the figures represents
the approximate trend. So the approximate angular speeds within the
corresponding interval can be evaluated. Figure
\ref{trend_clockwise_anormal} -- \ref{preceding_north} present the
rotations of sample complex sunspot groups ($\beta$, $\gamma$,
$\delta$, $\beta\gamma$,  $\beta\gamma\delta$ class). Figure
\ref{trend_clockwise_anormal} shows that the following sunspots with negative
magnetic polarity rotate clockwise. These sunspots are located in the
southern hemisphere. In the northern hemisphere, the following sunspots
with positive polarity rotate counterclockwise, which is shown in Figure
\ref{trend_anormal}. The examples of preceding sunspots are shown in
Figure \ref{preceding_south} and Figure \ref{preceding_north}.

 \begin{figure}
 \centering
  \includegraphics[width=0.9\textwidth]{trend.eps}
  \caption{10 $\alpha$ sunspot groups (from (a) to (j)) are located
           in the southern hemisphere in the year of 2014/2015. Their magnetic
           polarity is positive. They rotate counterclockwise, and the tendency
           is marked as red dashed lines. The X-axis refers to the observational
           time period of these sunspots, and the Y-axis refers to the
           $\theta$ coordinate in degrees.}
  \label{trend}
 \end{figure}

\begin{figure}
 \centering
  \includegraphics[width=0.9\textwidth]{preceding_north.eps}
  \caption{2 preceding sunspots of 2014 located in the northern
           hemisphere rotate clockwise, and their polarity is negative.
           The trend is the same as the $\alpha$ sunspot groups
           in the same hemisphere.}
  \label{preceding_north}
 \end{figure}


Table \ref{tb1} lists the characteristics of these sunspots. It shows
that there is a certain relationship between rotational direction and
magnetic polarity for sunspots. For the $\alpha$ sunspot groups,
sunspots
located in the southern hemisphere rotate counterclockwise,
and their magnetic polarities are positive.
The sunspots located in the northern hemisphere rotate clockwise, and their
magnetic polarities are negative. Closer to umbrae, we can obtain the
main rotational trends more easily. We list the radii we choose
to estimate the angular speeds.

\clearpage
\begin{deluxetable}{cclccc}
\tabletypesize{\scriptsize}
  \tablecaption{Sunspots in 2014/2015 \label{tb1}}
\tablewidth{0pt}
\tablehead{
\colhead{Number} & \colhead{NOAA AR} & \colhead{Location} & \colhead{Hale Class} & \colhead{Angular Speed ($\mathrm degree\ hr^{-1}$)} &
\colhead{Magnetic Polarity (Maximum Field)}
}
\startdata
  1 & 11948 & N06 & $\alpha/\alpha$ & +1.77 (4$''$.5) & N (-2144.60 Gs)\\
  2 & 12005 & N13 & $\alpha/\alpha$ & +1.46 (6$''$.0) & N (-2384.30 Gs)\\
  3 & 12022 & N17 & $\alpha/\alpha$ & +1.25 (2$''$.5) & N (-1333.50 Gs)\\
  4 & 12033 & N12 & $\alpha/\alpha$ & +1.42 (15$''$.0) & N (-1882.80 Gs)\\
  5 & 12050 & N12 & $\alpha/\alpha$ & +1.67 (2$''$.5) & N (-1674.40 Gs)\\
  6 & 12079 & N12 & $\alpha/\alpha$ & +1.04 (7$''$.5) & N (-2265.90 Gs)\\
  7 & 12218 & N16 & $\alpha/\alpha$ & +1.25 (8$''$.0) & N (-2271.30 Gs)\\
  8 & 12272 & N12 & $\alpha/\alpha$ & +2.67 (3$''$.0) & N (-1338.50 Gs)\\
  9 & 11973 & N06 (preceding) & $\beta/\beta$ & +1.46 (4$''$.5) & N (-1829.20 Gs)\\
  10 & 12017 & N09 (preceding) & $\beta/\beta$ & +1.25 (6$''$.5) & N (-1816.10 Gs)\\
    \hline
  1 & 11946 & N09 (following) & $\beta\gamma/\beta$ & -2.58 (18$''$.5) & P (2098.60 Gs)\\
  2 & 11968 & N10 (following) & $\beta\gamma/\beta\gamma$ & -1.95 (8$''$.5) & P (1851.80 Gs)\\
    \hline
    \hline
  1 & 11949 & S16 & $\alpha/\alpha$ & -1.46 (5$''$.5) & P (2034.90 Gs)\\
  2 & 11955 & S14 & $\alpha/\alpha$ & -1.67 (3$''$.5) & P (1788.20 Gs)\\
  3 & 12061 & S24 & $\alpha/\alpha$ & -1.04 (8$''$.5) & P (1685.50 Gs)\\
  4 & 12075 & S09 & $\alpha/\alpha$ & -2.08 (4$''$.5) & P (1821.80 Gs)\\
  5 & 12151 & S08 & $\alpha/\alpha$ & -1.35 (7$''$.0) & P (2239.10 Gs)\\
  6 & 12194 & S12 & $\alpha/\alpha$ & -0.94 (4$''$.0) & P (1851.10 Gs)\\
  7 & 12200 & S16 & $\alpha/\alpha$ & -1.80 (3$''$.0) & P (1373.10 Gs)\\
  8 & 12227 & S04 & $\alpha/\alpha$ & -0.73 (5$''$.5) & P (2079.40 Gs)\\
  9 & 12252 & S20 & $\alpha/\alpha$ & -1.77 (4$''$.0) & P (1792.60 Gs)\\
  10 & 12261 & S11 & $\alpha/\alpha$ & -1.04 (4$''$.5) & P (1772.90 Gs)\\
  11 & 11974 & S13 (preceding) & $\beta\gamma/\beta\gamma$ & -0.43 (9$''$.0) & P (2062.40 Gs)\\
  12 & 11977 & S10 (preceding) & $\beta\gamma/\beta\gamma$ & -0.29 (18$''$.5) & P (1921.60 Gs)\\
  13 & 11991 & S24 (preceding) & $\beta\gamma/\beta\gamma$ & -0.52 (4$''$.5) & P (1889.60 Gs)\\
  14 & 12002 & S18 (preceding) & $\beta\gamma\delta/\beta\gamma\delta$ & -1.60 (5$''$.5) & P (2001.20 Gs)\\
   \hline
  1 & 11967 & S13 (following) & $\beta\gamma\delta/\beta\gamma\delta$ & +1.84 (19$''$.0) & N (-2660.40 Gs)\\
  2 & 11974 & S13 (following) & $\beta\gamma/\beta\gamma$ & +2.94 (6$''$.0) & N (-1645.20 Gs)\\
  3 & 11977 & S10 (following) & $\beta\gamma/\beta\gamma$ & +1.73 (5$''$.0) & N (-1946.40 Gs)\\
  4 & 11982 & S10 (following) & $\beta\gamma/\beta\gamma$ & +1.14 (15$''$.0) & N (-2458.30 Gs)\\
  5 & 11991 & S24 (following) & $\beta\gamma/\beta\gamma$ & +2.62 (11$''$.5) & N (-2129.40 Gs)\\
  6 & 12002 & S18 (following) & $\beta\gamma\delta/\beta\gamma\delta$ & +0.63 (10$''$.5) & N (-1630.70 Gs)\\
  7 & 12010 & S15 (following) & $\beta\gamma/\beta\gamma$ & +2.43 (8$''$.0) & N (-1822.70 Gs)\\
  8 & 12049 & S07 (following) & $\beta\gamma/\beta\gamma$ & +1.00 (8$''$.5) & N (-1832.30 Gs)\\
   \enddata
   \tablecomments{Data are from HMI. In the ``location" column, ``N" stands for northern
                  hemisphere and ``S" stands for southern hemisphere;
                  in the ``Angular Speed" column, ``+" stands for rotating
                  clockwise  and ``-" stands for rotating counterclockwise;
                  in the ``Magnetic Polarity" column, ``P" stands for
                  positive and ``N" stands for ``negative".}
\end{deluxetable}


The complex active regions evolve fast and have two or more sunspots
with different magnetic polarities. We choose the sunspot with singular
polarity to analyze the rotation. The results are also listed in Table
\ref{tb1}. The polarities and rotational directions of the
preceding sunspots are the same as the $\alpha$ sunspot groups located in
the same hemisphere. But sunspots with negative polarity located in the
southern hemisphere (the following sunspots) rotate clockwise in general,
which is contrary to the $\alpha$ sunspot groups located in the same
hemisphere. Similarly, the trend of sunspots located in the northern
hemisphere is also contrary to the northern hemispheric $\alpha$ sunspot
groups. They rotate counterclockwise while their magnetic polarities are
positive.

\begin{figure}
 \centering
  \includegraphics[width=0.9\textwidth]{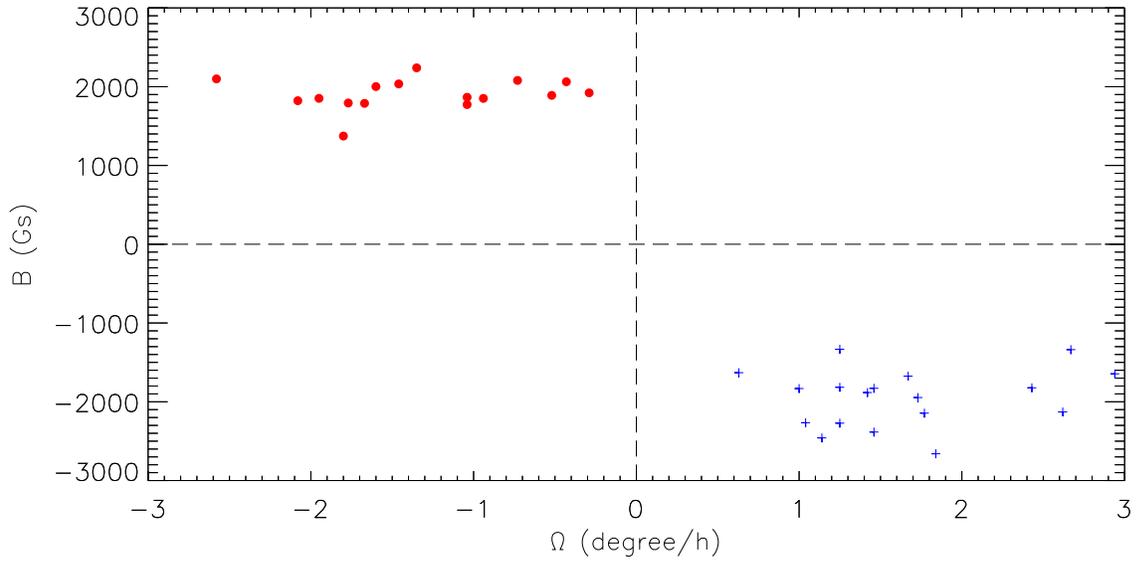}
  \caption{Magnetic field versus angular speed $\Omega$ of the
           rotating sunspots in 2014/2015. The rotational direction is related
           to the magnetic polarity. The sunspots marked with red filled
           circles rotate counterclockwise, and they have positive magnetic
           polarity. The sunspots marked with blue plus signs rotate clockwise and
           they have negative magnetic polarity.}
  \label{wb1}
 \end{figure}

We plot the maximum of magnetic field strengths versus angular speeds
of all sunspots in Figure \ref{wb1}. It shows clearly
that sunspots with positive polarity rotate counterclockwise, and
sunspots with negative polarity rotate clockwise.

\subsection{The Cycle of Sunspots' Rotation}\label{cycle}
In order to seek probable periodicity of rotation, we study sunspot data (the
$\alpha$ sunspot groups and the preceding sunspots) of 2 yr in
two different solar cycles with the method described in section \ref{method}.
The MDI data of 2003 in solar cycle 23 (from 1997 to 2008) and HMI
data of 2014/2015 in solar cycle 24 (from 2008 to 2019) are used in our study.

\begin{figure}
 \centering
  \includegraphics[width=0.9\textwidth]{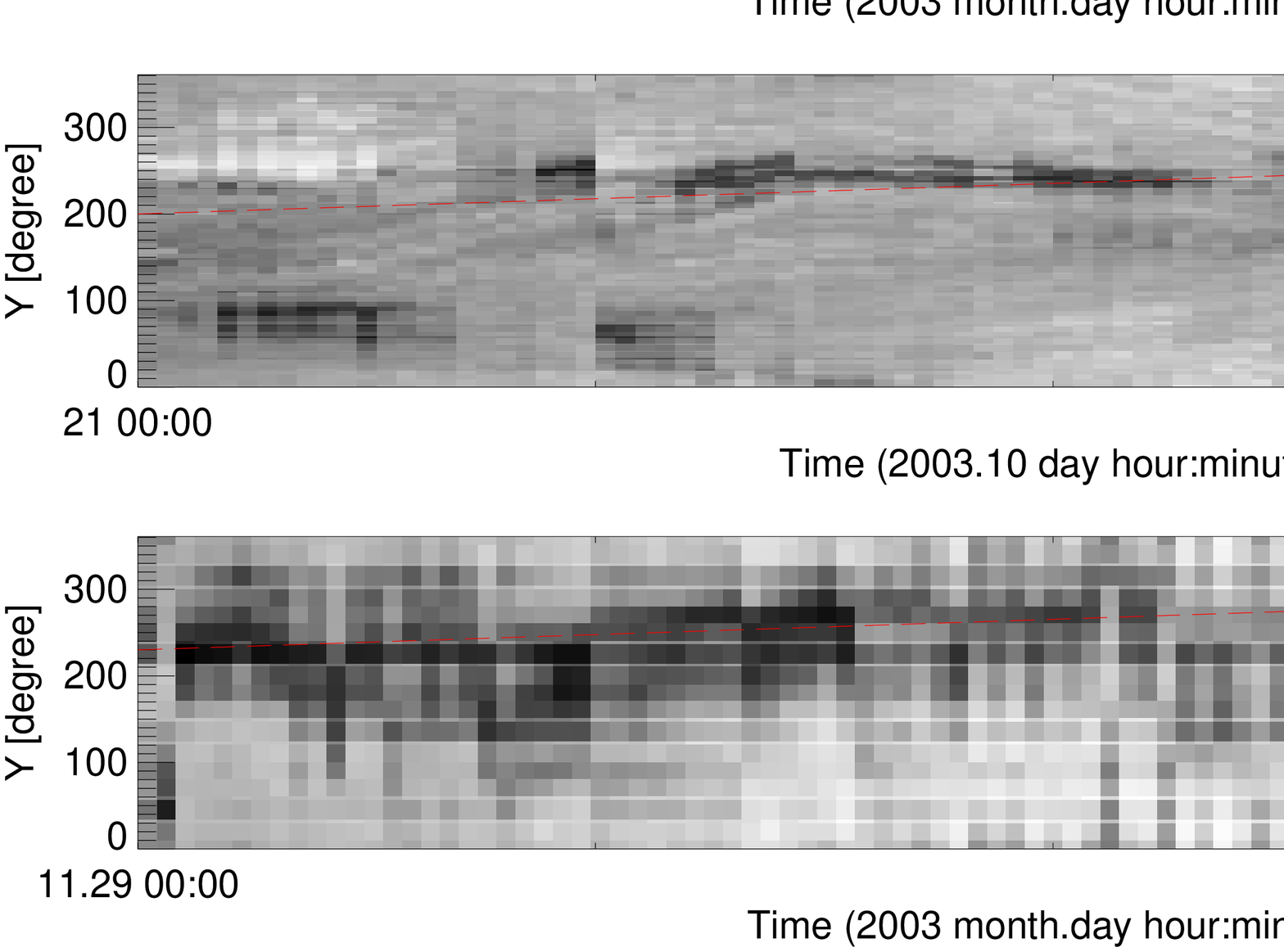}
  \caption{The 22 $\alpha$ sunspot groups and preceding sunspots of
           2003 located in the northern hemisphere rotate counterclockwise,
           and their polarity is positive.}
  \label{mdi2003_north}
\end{figure}
\begin{figure}
 \centering
  \includegraphics[width=0.9\textwidth]{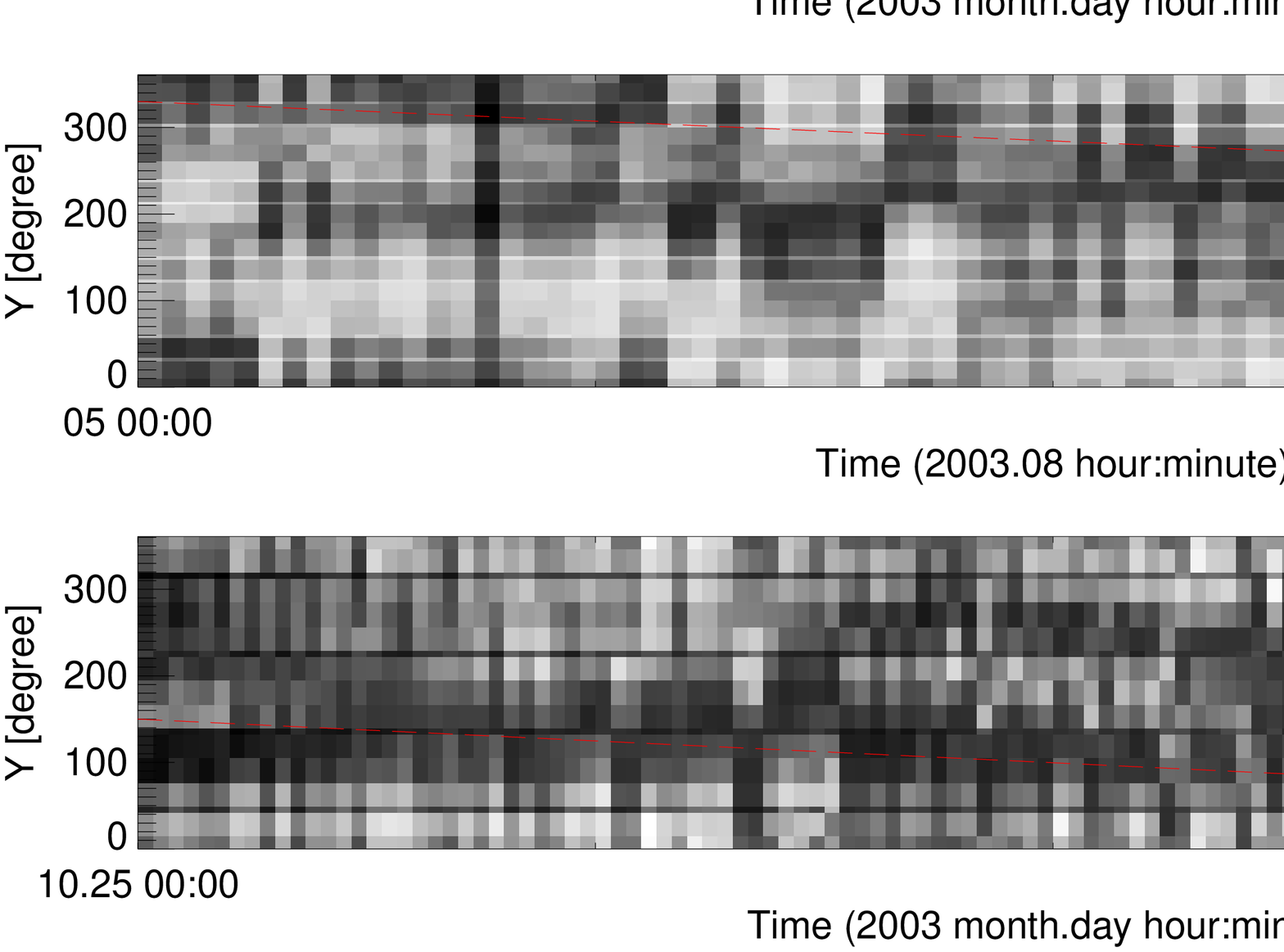}
  \caption{The 20 $\alpha$ sunspot groups and preceding sunspots of
           2003 located in the southern hemisphere rotate clockwise,
           and their polarity is negative.}
  \label{mdi2003_south}
 \end{figure}

There are 42 sunspots in 2003 (22 of them are located in the northern
hemisphere, and 20 of them are located in the southern hemisphere) in our
sample. Figure \ref{mdi2003_north} shows the rotation of the $\alpha$ sunspot
groups and the preceding sunspots located in the northern hemisphere. These sunspots
rotate counterclockwise and have positive polarity. The 20 sunspots located
in the southern hemisphere rotate clockwise, as shown in Figure \ref{mdi2003_south}.
Their magnetic polarities are negative.
We obtain the
angular speeds, locations, and magnetic strengths of these sunspots as shown
in the tables. Table \ref{tb2} lists the the sunspots in 2003 (cycle 23).
The rotational directions of these sunspots have hemispheric
preference: sunspots in the northern hemisphere rotate counterclockwise,
and sunspots in the southern hemisphere rotate clockwise.

\clearpage
\begin{deluxetable}{cclccc}
\tabletypesize{\scriptsize}
\tablecaption{Sunspots in 2003 \label{tb2}}
\tablewidth{0pt}
\tablehead{
\colhead{Number} & \colhead{NOAA AR} & \colhead{Location} & \colhead{Hale Class} & \colhead{Angular Speed ($\mathrm degree\ hr^{-1}$)} & \colhead{Magnetic Polarity (Maximum Field)}
}
\startdata
1 & 10258 & N07 (preceding) & $\alpha/\beta$ & -0.18 & P (2535.67 Gs)\\
2 & 10288 & N13 (preceding) & $\beta/\alpha$ & -0.98 & p (2244.02 Gs)\\
3 & 10290 & N17 (preceding) & $\beta\gamma/\beta\gamma$ & -0.48 & P (2464.84 Gs)\\
4 & 10296 & N12 (preceding) & $\beta\gamma/\beta\gamma$ & -0.30 & P (3214.95 Gs)\\
5 & 10306 & N07 (preceding) & $\beta\gamma/\beta$ & -0.21 & P (3227.82 Gs)\\
6 & 10319 & N13 (preceding) & $\beta/\beta$ & -0.92 & P (2771.54 Gs)\\
7 & 10325 & N10 (preceding) & $\beta/\beta$ & -0.31 & P (3038.31 Gs)\\
8 & 10330 & N07 (preceding) & $\beta\gamma/\beta$ & -0.35 & P (3070.89 Gs)\\
9 & 10346 & N16 (preceding) & $\beta/\alpha$ & -0.35 & P (2687.02 Gs)\\
10 & 10351 & N08 & $\alpha/\alpha$ & -0.12 & P (2622.57 Gs)\\
11 & 10373 & N07 (preceding) & $\beta/\alpha$ & -0.76 & P (2715.43 Gs)\\
12 & 10375 & N12 (preceding) & $\beta\gamma\delta/\beta\gamma$ & -1.05 & P (2733.37 Gs)\\
13 & 10377 & N04 (preceding) & $\beta/\beta$ & -0.56 & P (2180.81 Gs)\\
14 & 10387 & N18 (preceding) & $\beta\gamma/\beta\gamma$ & -0.62 & P (2425.00 Gs)\\
15 & 10390 & N14 (preceding) & $\alpha/\beta$ & -1.23 & P (2501.51 Gs)\\
16 & 10420 & N11 & $\alpha/\alpha$ & -2.34 & P (2758.44 Gs)\\
17 & 10463 & N09 & $\alpha/\alpha$ & -0.72 & P (2428.32 Gs)\\
18 & 10465 & N00 (preceding) & $\beta/\beta$ & -0.21 & P (2593.09 Gs)\\
19 & 10484 & N04 (preceding) & $\beta\gamma\delta/\beta\gamma\delta$ & -0.49 & P (2472.54 Gs)\\
20 & 10488 & N09 (preceding) & $\beta/-$ & -1.28 & P (2869.35 Gs)\\
21 & 10513 & N13  & $\alpha/\alpha$ & -0.42 & P (2507.05 Gs)\\
22 & 10528 & N09 (preceding) & $\beta\gamma/\beta\gamma$ & -0.25 & P (2993.30 Gs)\\
    \hline
    \hline
1 & 10242 & S08 (preceding) & $\beta/\beta\gamma$ & +0.25 & N (-2433.70 Gs)\\
2 & 10244 & S23 (preceding) & $\beta\gamma/\beta$ & +0.59 & N (-2141.60 Gs)\\
3 & 10250 & S26 (preceding) & $\beta/\beta$ & +0.35 & N (-2539.46 Gs)\\
4 & 10254 & S15 (preceding) & $\beta/\beta$ & +0.63 & N (-2548.26 Gs)\\
5 & 10255 & S13 (preceding) & $\beta/\beta$ & +0.97 & N (-2823.57 Gs)\\
6 & 10281 & S15 & $\alpha/\alpha$ & +0.058 & N (-1947.11 Gs)\\
7 & 10314 & S14 (preceding) & $\beta\gamma/-$ & +2.44 & N (-2621.95 Gs)\\
8 & 10334 & S08 (preceding) & $\beta/\beta$ & +0.28 & N (-2508.12 Gs)\\
9 & 10337 & S04 (preceding) & $\beta\gamma/\beta\gamma$ & +0.71 & N (-2678.66 Gs)\\
10 & 10345 & S16 (preceding) & $\beta/\beta$ & +1.39 & N (-2427.95 Gs)\\
11 & 10349 & S13 (preceding) & $\beta\gamma/\beta\gamma$ & +0.99 & N (-3176.53 Gs)\\
12 & 10380 & S15 (preceding) & $\beta\gamma/\beta\gamma\delta$ & +0.088 & N (-2802.09 Gs)\\
13 & 10381 & S18 (preceding) & $\beta/\beta$ & +0.24 & N (-2128.26 Gs)\\
14 & 10386 & S07 (preceding) & $\beta\gamma\delta/\beta\gamma\delta$ & +0.48 & N (-1755.27 Gs)\\
15 & 10421 & S08 (preceding) & $\beta/\beta$ & +0.26 & N (-2696.65 Gs)\\
16 & 10424 & S18 (preceding) & $\beta\gamma\delta/\beta\gamma$ & +0.25 & N (-2789.32 Gs)\\
17 & 10425 & S09 (preceding) & $\beta/\beta$ & +0.75 & N (-2623.42 Gs)\\
18 & 10473 & S09 (preceding) & $\beta/\alpha$ & +0.39 & N (-2632.23 Gs)\\
19 & 10486 & S16 (preceding) & $\beta\gamma\delta/\beta\gamma\delta$ & +0.52 & N (-2701.67 Gs)\\
20 & 10495 & S22 (preceding) & $\beta/-$ & +1.54 & N (-2237.67 Gs)\\
\enddata
\tablecomments{Data from MDI. In the ``location" column, ``N" stands for northern
                  hemisphere and ``S" stands for southern hemisphere;
                  in the ``Angular Speed" column, ``+" stands for rotating
                  clockwise  and ``-" stands for rotating counterclockwise;
                  in ``Magnetic Polarity" column, ``P" stands for
                  positive and ``N" stands for negative.}
\end{deluxetable} 


The polarities reverse their
sense 11 yr later, as listed in Table \ref{tb1}. These sunspots
were observed in 2014/2015 (cycle 24), which contain 24 sunspot
groups (we just
focus on the $\alpha$ sunspot groups and the preceding sunspots), 10 of which
are located in the northern hemisphere and 14 of which are located in the
southern hemisphere. According to the study of section \ref{relation} sunspots
in the northern hemisphere rotate clockwise, and sunspots in the
southern hemisphere rotate counterclockwise.


We can see from the Table \ref{tb1} and \ref{tb2} that all the
$\alpha$ sunspot groups and preceding sunspots with positive
polarity rotate counterclockwise and sunspots with negative polarity
rotate clockwise.

\begin{figure}
 \centering
  \includegraphics[width=0.9\textwidth]{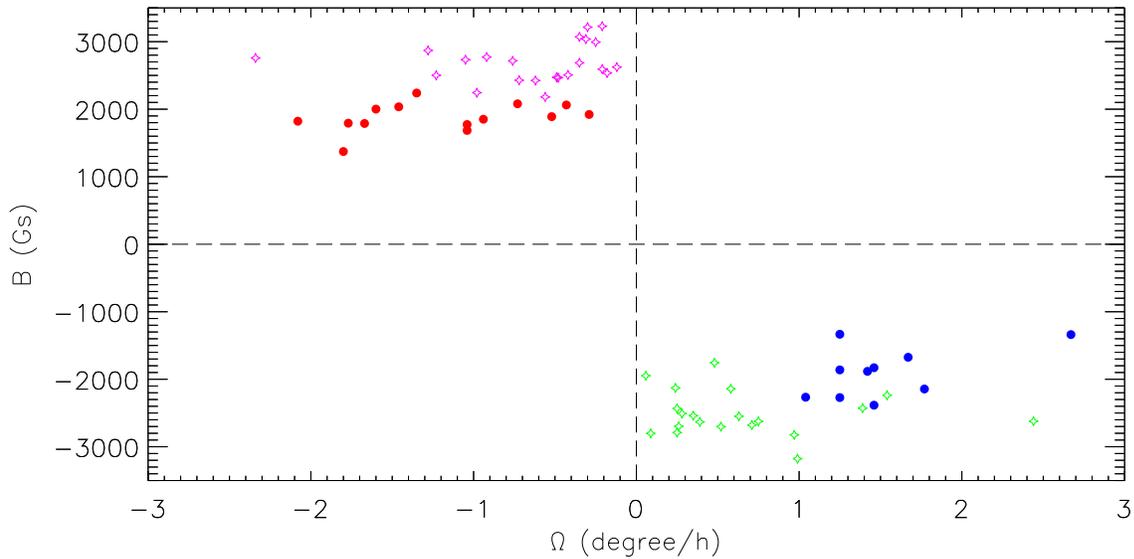}
  \caption{Sunspots' magnetic field versus angular speed $\Omega$. The
           star symbols represent the sunspots in the year of 2003,
           and the filled circle symbols represent the sunspots in the year
           of 2014/2015.}
  \label{wb2}
 \end{figure}

Figure \ref{wb2} shows the scatter diagram of the sunspots. It is a plot of the
the maximum of magnetic field strengths versus angular speeds of sunspots.
The star represent the sunspots in the year of
2003, and the filled circle represent the sunspots in the year
of 2014/2015. The range of the speed value in 2014 is from about
$\sim 0^{\circ}.29 \rm \ h^{-1}$ to $2^{\circ}.67 \rm \ h^{-1}$, and in 2003 it
is from about $\sim 0.^{\circ}058 \rm \ h^{-1}$ to $2^{\circ}.44 \rm \ h^{-1}$.
Because of the lower spatial resolution of MDI (4$''$;
\citealt{sch1995}), we cannot follow some fine rotational features.
However, the resolution of HMI is better (0$''$.5; \citealt{sch2012}),
which allows us to detect the rotational features more easily.
Figure \ref{wb2} reveals that the sunspots with positive
polarity rotate counterclockwise and sunspots with negative polarity
rotate clockwise in both solar cycles. The rotational direction
corresponds to the magnetic polarity independent of the solar cycle.

\begin{figure}
 \centering
  \includegraphics[width=0.9\textwidth]{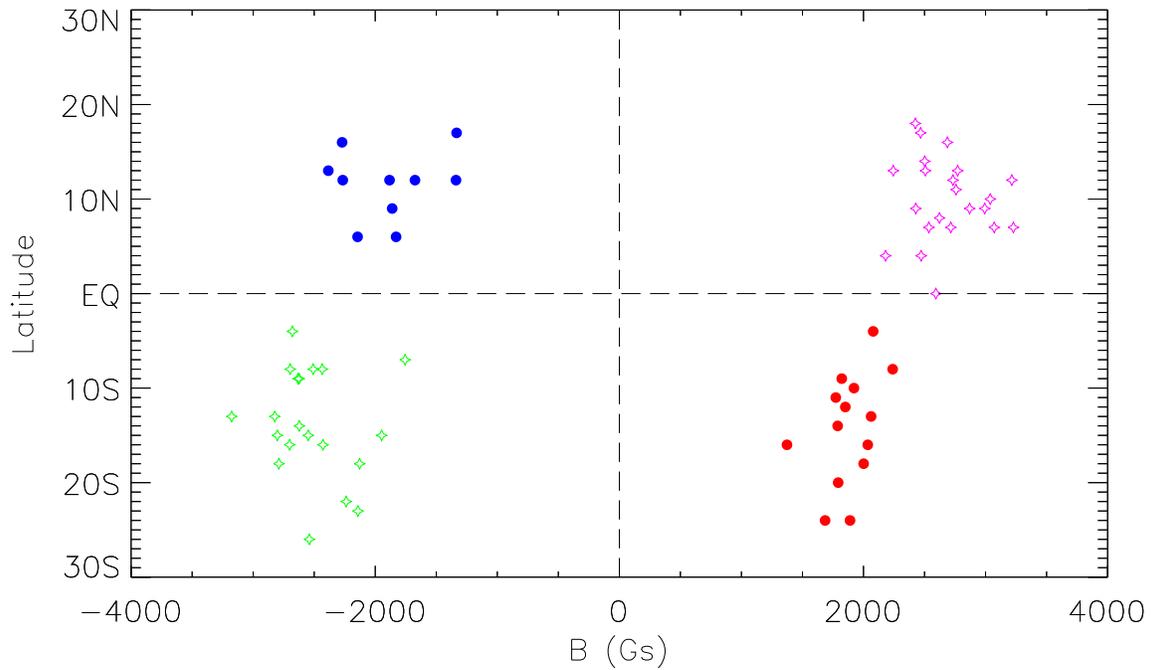}
  \caption{Maximum of magnetic field strengths of the $\alpha$ sunspots
           and the preceding sunspots in both hemispheres. The star
           symbols represent the sunspots in the year of 2003, and the
           filled circle symbols represent the sunspots in the year of 2014/2015.
           The sunspots with positive polarity rotate counterclockwise,
           and the sunspots with negative polarity rotate clockwise.}
  \label{lb}
 \end{figure}

The $\alpha$ sunspot groups and the preceding
sunspots in the northern and southern hemispheres have opposite
polarities (Hale's law), which is shown in Figure \ref{lb}. In 2003,
the $\alpha$ sunspot groups and the preceding sunspots have positive
polarity in the northern hemisphere and negative
polarity in the southern hemisphere (star symbols), but the
situation is reversed in the year of 2014/2015
(filled circle symbols).
The polarity of the $\alpha$ sunspot groups and the preceding sunspots
in each hemisphere reverses the sense every 11 yr.

\begin{figure}
 \centering
  \includegraphics[width=0.9\textwidth]{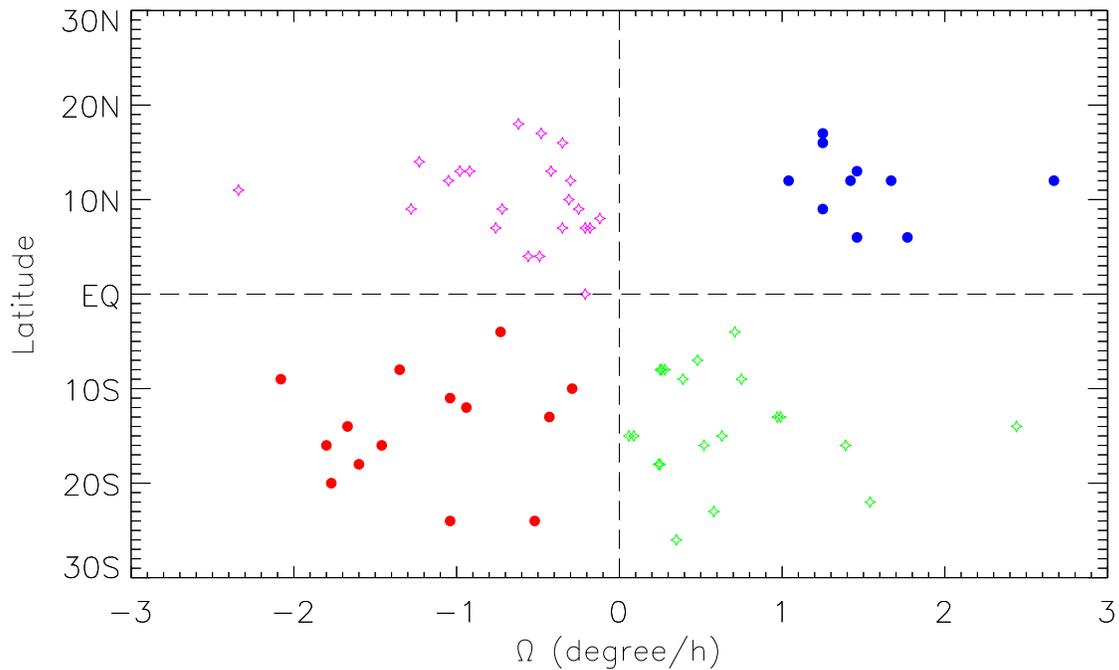}
  \caption{Scatter of sunspots about the latitude and angular speed. The
           star symbols represent the sunspots in the year of 2003, and the
           filled circle symbols represent the sunspots in the year of 2014/2015.}
  \label{wl}
 \end{figure}

The polarity of sunspots reverses its sense every 11 yr solar cycle, and that
makes the 22 yr period magnetic (Hale) cycle. The rotational directions
and magnetic polarities have a certain relationship, so the
change of the rotational directions seems to have the same pattern as the cycle. The
angular speed of the $\alpha$ sunspot groups and preceding sunspots is
shown in Figure \ref{wl}. In the year of 2003, the $\alpha$
sunspot groups and the preceding sunspots rotate counterclockwise
in the northern hemisphere and rotate clockwise in the southern hemisphere.
While from 2014 January to 2015 February, the $\alpha$ sunspot groups and the
preceding sunspots rotate clockwise in the northern hemisphere
and rotate counterclockwise in the southern hemisphere.
So rotational directions of $\alpha$ sunspot groups and preceding sunspots
change with magnetic polarities (Hale Cycle). In the northern hemisphere,
sunspots in 2003 of the Cycle 23 rotate counterclockwise and sunspots in
2014/2015
of the Cycle 24 reverse their rotational direction (clockwise). In the southern
hemisphere, the sunspots in different solar cycles also have opposite
rotational directions.

\section{Summary and Discussion}\label{summary}
The relatively large and stable sunspots rotate with a variety of angular velocities.
Some sunspots rotate obviously, and the rotational features
can be followed or seen from the movies easily and qualitatively.
However, for some
sunspots, the features are not so obvious, and their rotational trends
cannot be obtained easily. Thus, sunspots are uncurled from the Cartesian
frame to the polar frame. Then the radius of the uncurled circle is
fixed to obtain time slices. The trends of time slices
indicate the rotational directions, and the ranges of the trends are
rotational magnitudes in degrees. In our study, we focus on the rotational
directions of sunspots. The rotations of the $\alpha$ sunspot groups
have continuity as structures evolving smoothly. The specific features of
the $\alpha$ sunspot groups have enough lifetime to be traced.
By calculating from the trends of the time slices,
the angular speeds of about $\pm1^{\circ}.31\rm\ hr^{-1}$ are found (the $\alpha$
sunspots rotate a little more than $100^{\circ}$ within 4 days).
The rotations of the sunspots in complex active regions are more
complicated. For example, we can see
from the movie that the sunspots rotate clockwise; however, in the time-slice
figures, there are not only streaks indicating clockwise rotation but also
streaks indicating counterclockwise rotation. This makes it difficult to determine
their rotational directions. Besides, uniform
rotation of the whole sunspot does not last for a long time. So in our
sample, we try to choose the relatively large and stable sunspots that
have relatively few eruptive events in these regions. The rotational
features have enough lifetime to allow us to trace. After calculating, the mean
angular speed is about $\pm0.91^{\circ}\rm\ hr^{-1}$.


From the statistical analysis of rotating sunspots in the year of
2014/2015 in section \ref{relation}, we find the following relations
between rotation and magnetic polarities:


i. The $\alpha$ sunspot groups. These sunspots tend to rotate
counterclockwise and have positive magnetic polarity in the southern
hemisphere, whereas in the northern hemisphere, the $\alpha$ sunspots
tend to rotate clockwise and have negative magnetic polarity.


ii. The preceding sunspots in the complex sunspot groups. These
sunspots and the $\alpha$ sunspot groups
show the same tendency between magnetic polarities
and rotational directions. In the southern hemisphere, they tend to rotate
counterclockwise and have positive polarity. The results are
opposite in the northern hemisphere. But those sunspots evolve
in a much more complicated pattern than the $\alpha$ sunspot groups.


iii. The following sunspots in the complex sunspot groups.
These complex sunspots have
negative magnetic polarity and tend to rotate clockwise in the southern
hemisphere, whereas in the northern hemisphere,
the sunspots have positive magnetic polarity and tend to rotate counterclockwise.


According to the above results, in our sample the sunspots
corresponding to negative magnetic polarity tend to rotate clockwise,
and sunspots corresponding to positive magnetic polarity tend to rotate
counterclockwise, no matter they are in the northern or southern hemisphere.


With the data from two solar cycles, we find that the rotational direction of
sunspots follows periodicity. It is the same as magnetic polarity of sunspots.
In this study, we choose the sunspot data of two solar cycles from
MDI/\it SOHO \rm and HMI/\it SDO\rm, respectively. In the year of 2003, 5 $\alpha$
sunspot groups and 37 preceding sunspots of complex sunspot groups are
studied. There are 22 sunspots located in the northern hemisphere and 20
sunspots located in the southern hemisphere. In the northern hemisphere,
the sunspots have positive polarity. They are rotating
counterclockwise. The 11 sunspots in the southern hemisphere rotate
clockwise and have negative polarity. The results agree with the
conclusion that sunspots with positive polarity rotate counterclockwise
and sunspots with negative polarity rotate clockwise.


From 2014 January to 2015 February, there are 10 $\alpha$
sunspot groups and 4 preceding
sunspots with positive polarity rotating counterclockwise in the southern
hemisphere. In the northern hemisphere, 8 $\alpha$ sunspot groups and 2
preceding sunspots with negative polarity rotate clockwise.


We study sunspots of 2 yr in two successive solar cycles.
The polarities of the $\alpha$ sunspot groups and preceding
sunspots in both hemispheres reverse in these 2 yr.
This is the result of the 22 yr magnetic cycle of sunspots.
The directions of sunspots' rotation are also reversed.


It is recognized that the magnetic field of sunspots
is connected to the solar dynamo, but it is still not well understood.
Through the study on the rotation of sunspots, we find
the correlation of rotation and magnetic polarity of
sunspots. This may help
us to understand the solar dynamo in the future studies.

\acknowledgments
We thank the referee for constructive suggestions and comments
that helped to improve this paper.
The authors thank the \it SDO\rm /HMI and \it SOHO\rm /MDI team for providing the data.
This work is supported by the National Natural Science Foundations of
China (U1231104). H.W. acknowledges NSF AGS-1348513 and AGS-1408703.

\bibliography{ref}

\end{document}